\documentstyle[12pt]{article}
\def\vs{\vspace}
\def\vss{\vspace{.5cm}}

\def\begc{\begin{center}}
\def\endc{\end{center}}
\def\be{\begin{eqnarray}}
\def\ee{\end{eqnarray}}
\def\la{\label}
\def\rarw{\rightarrow}
\def\bfr{\begin{flushright}}
\def\efr{\end{flushright}}

\def\nn{\nonumber  \\}

\def\begv{\begin{verbatim}}
\def\endv{\end{verbatim}}
\def\a{\alpha}
\def\b{\beta}
\def\g{\gamma}

\def\d{\delta}

\def\h{\eta}
\def\th{\theta}

\def\k{\kappa}
\def\l{\lambda}

\def\m{\mu}
\def\n{\nu}

\def\o{\o}

\def\P{\Pi}

\def\r{\rho}

\def\f{\phi}
\def\F{\Phi}
\def\vf{\varphi}

\def\Ps{\Psi}

\def\bra{\langle}
\def\ket{\rangle}
\def\fr{\frac}
\def\del{\partial}
\def\mxxb{\left( \begin{array}{cc}}          
\def\mxxe{\end{array} \right)}
\begin{document}
\baselineskip=20pt
\renewcommand{\thefootnote}{\fnsymbol{footnote}}
\bfr
hep-th/9806111\\
TEP-13\\
June, 1998
\efr
$ $\vs{2cm}
\begc
\begin{Huge}
\begin{bf}
The Problem of Time in\vs{0.3cm}\\
the Quantum Theory of\vs{0.6cm}\\
Black Holes\vs{1cm}
\end{bf}
\end{Huge}
 \vs{1cm}

\begin{large}
Takayuki Hori\footnote{E-mail: hori@main.teikyo-u.ac.jp}
\vss

\begin{it}
Teikyo University, Hachioji,
  192-0395 Japan\end{it} \vs{1cm}
\end{large}

 \vss

\endc
We discuss the problem of time in spherically symmetric pure Einstein gravity 
with the cosmological term by using an exact solution to the Wheeler-DeWitt 
equation. 
A positive definite inner product is defined, based on the momentum constraint 
rather than the hamiltonian constraint.
A natural notion of time is introduced via the Heisenberg equation. This 
notion  enables one to reproduce the time-time component of the classical metric.
Non-Hermiticity of the hamiltonian is essential in the definition of time.\vss
\newpage
\pagestyle{plain}
\setcounter{page}{1}
The canonical theory of the general relativity reveals the complete constraint 
structure of the  theory\cite{Dirac}\cite{ADM}. 
Replacing the canonical momenta  by functional derivatives with respect to the 
corresponding fields, we obtain the Wheeler-DeWitt (WDW) equation 
\cite{DeWitt}\cite{Wheeler} from the classical constraints.
A long-standing problem in quantum cosmology is how to interpret the wave 
functional in the WDW theory.

A major problem results from the fact that the total hamiltonians of any 
models with general coordinate invariance  vanish. Thus any observables in 
such a theory are absolutely static with respect to the time coordinate.  
This is referred as the ``problem of time''(see, for example, 
Ref.~\cite{Kuchar}).
In the classical theory a surface term can be added to the hamiltonian (see, 
for example, Ref.~\cite{gegen}), but this violates the general coordinate 
invariance which must be preserved in the full quantum theory.
This prevents us from comparing the classical theory with the full quantum 
theory.
In the semiclassical approximation, the problem may be viewed as superficial, 
since one can define a reference clock of a semiclassical nature, and the WDW 
equation reduces to a Schr\"odinger-type equation of first order with respect 
to differentiation by the clock variable\cite{Banks}, while the latter is 
related to the coordinate time via a classical equation of motion.
However,  the genuine problem in the full quantum theory is that one should 
{\it derive}, in some limits, the classical metric  from the quantum theory 
itself.

Another problem of the WDW theory, referred to as the ``Hilbert space 
problem'',  concerns  the hyperbolic signature of the DeWitt metric 
\cite{DeWitt} in the superspace, which may lead to   negative probability, 
as in the case of the Klein-Gordon wave equation.
This has also been treated in a semiclassical approximation \cite{Vilenkin} 
and  in  minisuperspace approaches, and interesting results and concepts such 
as the third quantization have been proposed\cite{Rubakov}.

In the present paper we consider the above two problems in  spherically 
symmetric Einstein gravity with the cosmological term using  an exact 
solution to the WDW equation. An exact solution in the effectively  
two-dimensional gravity theories  was  first obtained in \cite{Henneaux} 
for  Jackiw-Teitelboim gravity\cite{Jackiw}, and in using similar method 
in \cite{Gegenberg} for  spherically symmetric gravity without matter fields.
Exact WDW solutions for the CGHS model\cite{CGHS}, a two-dimensional gravity 
with a dilaton, were obtained in Ref.~\cite{hori} in the case with conformal 
matter fields, and in Ref.\cite{Gegenberg2} in the case without matter fields.
In order to analyze the effect of Hawking radiation one needs matter fields.
Unfortunately no exact solution has been found in the spherically symmetric 
theory with matter fields due to their coupling to the scale factor (dilaton 
field).
Here we restrict ourselves to the two problems mentioned above and  apply the 
solution obtained in \cite{hori} to the spherically symmetric theory, 
neglecting the matter fields.


Let us start with the spherically symmetric world line element written as
\be
        ds^2  = -g_{\m\n}(t, r)dx^{\m}dx^{\n}  -  e^{-2\vf(t, r)}(d\th^2  + 
\sin^2{\th}d\f),
\la{metric}
\ee
where $e^{-2\vf}$ is a scale factor.
It turns out that the entire calculation is drastically simplified if one 
introduces the zweibein $e_{\m a}$ defined by $g_{\m\n} = 
e_{\m a}e_{\n}{}^{a}$
(we choose the signature of the flat metric as  $\h _{11} = -\h_{00} = 1$).
The Einstein action with the cosmological term is
\be
    S  &=&  \fr{1}{16\pi G}\int d^4x\ {}^{(4)}e(\ {}^{(4)}R - 2\l)  \nn
        &=&  \fr{1}{4G}\int d^2xe\left[e^{-2\vf}(R  +  
2g^{\m\n}\del_{\m}\vf\del_{\n}\vf  -  2\l)  + 2\right] ,
\la{action}
\ee
where $G$ is the gravitational constant and the superscript 4 designates the 
four-dimensional quantities.
The technical advantage of using the zweibein  variable is manifested in the 
canonical theory in the local Lorentz gauge defined by \cite{hori0}
\be
           e_{\m a}  = \mxxb     \a  &  \b e^{-\r}\\
                                 0  &  e^{\r}
                       \mxxe  ,
\la{LLgauge}
\ee
where $\a$ and $\b$ are the shift and the lapse functions, respectively, and 
$\r$ is another dynamical variable. 

The canonical theory of the spherically symmetric Einstein gravity was 
obtained long ago\cite{BCMN}, and the hamiltonian (with the cosmological term) 
is expressed as
\be
    H = \int dr(\b\F_1 + \a\F_2  + f_1\P_{\b} + f_2\P_{\a}),
\la{hamiltonian}
\ee
where $f_1 = \dot{\b}$ and $f_2 = \dot{\a}$ are not determined by the Legendre 
transformation and, hence, are arbitrary functions of the canonical variables.
$\F_1$ and $\F_2$ do not depend on $\a$ and $\b$, and are expressed in terms 
of the canonical momenta,
 $\P_{\r}, \P_{\vf}, \P_{\a}$ and $\P_{\b}$, conjugates of $\r, \vf, \a$ and 
$\b$, respectively, as
\def\delr{\fr{\del}{\del r}}
\be
        \F_1 &=& e^{-2\r}\left[\left( {\r}' - \delr \right)\P _{\r} + 
\vf '\P _{\vf}\right],\\
        \F_2 &=& \fr12 e^{-\r}\left[Ge^{2\vf}(\P_{\r} + 2\P _{\vf})\P_{\r}  
+   \fr{e^{-2\vf}}{G}R_1\right],\\
        R_1 &=& 3{\vf'}^2 - 2\vf ^{[2]} + \l e^{2\r} - 
e^{2\r  + 2\vf}.\la{R1}
\ee
Here the one-dimensional covariant derivatives \cite{hori} are defined by 
$\vf^{[0]} \equiv  \vf ,  \vf^{[n+1]} \equiv  (\delr  - n\r')\vf^{[n]} ~(n 
\ge  1)$ (primes denote derivatives with respect to $r$).
$R_1$ is the one-dimensional intrinsic curvature.

The WDW equation then becomes
\def\dr{\fr{\d}{\d \r(r)}}
\def\dfi{\fr{\d}{\d\vf(r)}}
\be
  \left[\left(\r '(r) - \delr\right)\dr  + \vf'(r)\dfi\right]\Ps{[}\r, \vf{]}  
= 0,
\la{kai}
\ee
\be
  \left[\left(\dr  + 2\dfi\right)\dr  - \fr{e^{-4\vf}}{G^2}R_1\right]\Ps{[}\r, 
\vf{]}  = 0.
\la{psai}
\ee
Because of the constraints, $\P_{\a} = \P_{\b} = 0$, the wave functional 
$\Ps$ cannot depend on $\a$ and $\b$.
Since the WDW equation contains second order functional differential 
operators  at the same point, it requires some regularization. 
In this paper we adopt  the DeWitt prescription\cite{DeWitt}, $\d(r - r) = 0$.
Furthermore, we set $\d^{(n)}(r - r) = 0  ~(n \ge 1)$.
The consistency of this regularization scheme will not be discussed in this 
paper, and it should be considered as one of the rules of the game.
There have been some attempts to derive the quantum effects of matter fields, 
in a semiclassical expansion for gravity, by setting $\d(0) \ne  0$ ~
\cite{Demers}.  This seems to be consistent only if one makes modifications 
of the WDW equation\cite{Cangemi}.
Another motive for our approach comes from the fact that our quantum theory  
is in fact an effective theory of the full four-dimensional theory rather 
than a genuine two-dimensional theory, where some Schwinger terms are 
unavoidable.

The key idea to solve the WDW equations is to introduce possible scalar and 
scalar densities out of $\r$ and $\vf$ in the one-dimensional sense
\cite{hori}.  That is,  we have two scalar densities, $\vf'$ and $e^{-\r}$, 
and one scalar $\vf$. Instead, it turns out to be convenient to use two 
scalars, $\vf$ and $X \equiv e^{-\r}\vf'$, and one scalar density $\vf'$. 
Since the momentum constraint Eq.~(\ref{kai}) amounts to  general covariance 
in the radial dimension, we are tempted to make an ansatz for the wave 
functional $\Ps$ as  
\be
\Ps{[}\r, \vf{]}  = \exp{\fr{1}{G}\int dre^{-2\vf}\vf'W(X, \vf)},
\la{ansatz}
\ee
where $W$ is a complex valued function of $X$, and $\vf$ is to be determined.
Then the momentum constraints Eq.~(\ref{kai}) reduce to the requirement that 
$W$ does not depend explicitly on $r$.
The hamiltonian constraint Eq.~(\ref{psai}) reduces to
\be
    (c_{\vf}  -  3c  - \l + e^{2\vf}) \vf'  + c_{X}X'  =  0,
\la{wdweq}
\ee
where $c  = X^2  +  X^4W_{X}^2$, 
and  $c_{\vf} = \del c/\del\vf, ~c_{X} = \del c/\del X$.
For Eq.~(\ref{wdweq}) to hold for an arbitrary function $\vf$, both 
coefficients of $\vf'$ {\it and} $X'$ must vanish. 
We see $c$ is a function of only $\vf$ because  $c_X = 0$.
Integrating $c  = X^2  +  X^4W_{X}^2$ with respect to $X$, and putting the 
coefficient of $\vf'$ in Eq.~(\ref{wdweq}) to zero, we obtain the solution.
So far we have used the variable $\r$, but in general $e^{2\r}$ becomes 
negative.
So we define $\g  = e^{2\r}$, which ranges from $-\infty$ to $+\infty$. 
The solution is then expressed as
\be
          W = \h^{1/2} +  i\ln\g^{-1/2}(\h^{1/2} -  i) +  i\ln(-\vf') + f(\vf),
\la{Wsol} 
\ee
where
\be
              \h  &=&  \fr{\g}{\g_c}  -  1, \la{defh}\\
              \g_c  &=&  \vf'^2e^{-2\vf}\left(1 - 2\k Ge^{\vf}  - 
\fr{\l}{3}e^{-2\vf}\right)^{-1},
\la{defgc}
\ee
and $\k$ is an arbitrary complex number with dimension of mass.
The multivalued terms in Eq.~(\ref{Wsol}) represent one of the branches 
in the complex plane, and we assume $\vf' < 0$ for simplicity.
The last term, $f(\vf)$, in Eq.~(\ref{Wsol}) is an arbitrary function of 
$\vf$, contributing to $\Ps$ only at the boundary and playing the role of 
suppressing  possible divergences.
The wave functional is stationary, both for $\g$ and $\vf$, at the function 
$\g = \g_c$, which coincides with $g_{11}$ in the Schwarzschild coordinate 
if one chooses $e^{-2\vf} = r^2$. 
The above solution satisfies such a boundary condition that if $\g(r)$ 
goes to $\infty$ for a fixed value of $r$ (except at the horizon) the wave 
functional vanishes, since the Hilbert space is assumed to be composed of 
functionals on the space of all well-behaved (continuous) functions of $r$. 

The inner product in the Hilbert space is defined by imposing its invariance 
with respect to a clock variable which we choose to be $\vf(r)$.
The ordinary definition due to DeWitt is, however, not positive definite.
Alternatively, we can define the  positive definite inner product as
\be
         \bra \Ps_1|\Ps_2\ket  = \int\prod_{r'}d\r(r')\Ps_1^*\Ps_2.
\la{inner1}
\ee
We can easily prove that Eq.~(\ref{inner1}) is independent of $\vf(r)$ due 
to the momentum constraint Eq.~(\ref{kai}), by which we have
\be
             \fr{\d}{\d\vf(r)}\Ps^*_1\Ps_2 = \fr{1}{\vf'(r)}\left(-\r'(r) + 
\fr{\del}{\del r}\right)\fr{\d}{\d\r(r)}\Ps^*_1\Ps_2.
\la{proof1}
\ee
According to the boundary condition imposed on $\Ps$, the functional integral 
of Eq.~(\ref{proof1}) over $\r$ vanishes (note  $\d'(0) = 0$).
The above argument can be applied equally to the four-dimensional case and 
to a model with matter fields as well.
The positive definiteness of the inner product defined by Eq.~(\ref{inner1}) 
allows us to interpret it as the probability amplitude.

The hamiltonian operator $H$ defined by Eq.~(\ref{hamiltonian}) with the 
replacement $\P_q \rarw  -i\d/\d q$ for a dynamical variable $q$, is 
{\it not} Hermitian in the present definition of the inner product, which 
contrasts with the situations considered by other authors (see, for example, 
Ref.~\cite{Gegenberg}).
However, there may be no physical reason for the Hermiticity of the 
hamiltonian constraint (recall the analogy to the relativistic particle).

Let us turn to the problem of obtaining the classical counterpart to the wave 
functional satisfying the WDW equation.
Since the functional integral of $(\d/\d\r)\Ps^*\Ps$ over $\r$ vanishes, we 
see that  
$\bra\h^{1/2}(r)\ket = 0$.
Similarly, we see that $\bra\h^{1/2}(r)\h^{1/2}(r')\ket = 0$ for $r \ne r'$. 
In the limit $r' \rarw r$ we obtain  $g_{11} =  \bra \g\ket  = \g_c$.
Choosing the ``gauge'' $e^{-2\vf} = r^2$, we get the Schwarzschild form of 
$g_{11}$,
\be
     g_{11} = \left( 1 - \fr{2\k G}{r} - \fr{\l}{3}r^2\right)^{-1}.
\ee
Thus $\k$ is interpreted as the mass located at origin.
Following reasoning similar to that above we see $\bra\h^{n/2}\ket = 0$ 
for an arbitrary integer $n$. This implies $\bra F(\g)\ket = F(\g_c)$ for 
an arbitrary function $F$.
For example, the one-dimensional intrinsic curvature,  $R_1  = -(3\vf'^2 - 
2\vf^{(2)})\h  + \vf'\h'$, has a vanishing expectation value.

The derivation of the time components of the metric is not straightforward, 
since the notion of ``time'' is implicit (or undefined) in the WDW 
quantization scheme.
Note that a state depends on the clock, which we choose to be $\vf(r)$, but 
does {\it not} depend on time.
Our formulation is in analogy with  ordinary quantum mechanics in the 
Heisenberg representation,  where an operator depends on time (not the clock) 
but a state does not.
In our formulation, however,  the time dependence is implicit in the field 
variables, and a priori there is no criterion to define the time dependence 
of any quantities. This is the ``problem of time'' mentioned above.
We assume that the time development of a quantity, ${\cal O}$, composed of 
$\r$ and $\vf$  and the functional derivatives with respect to them is 
{\it defined} by
\be
            \dot{\cal O} = i{[}H, {\cal O}{]}.
\la{Heisenberg}
\ee
For example functional derivatives with respect to the field variables become 
time dependent, and we denote them as ${\cal D}_q(r, t)$ instead of 
$\d/\d q(r)$ for $q = \f, \r$ (we assume ${\cal D}_q(r,  0) = \d /\d q(r)$).
This definition of time is meaningful only when the hamiltonian is {\it not} 
Hermitian, since otherwise we have $\bra\dot{{\cal O}}\ket = i\bra(H{\cal O} 
- {\cal O}H)\ket = i\bra(H^{\dag}{\cal O} - {\cal O}H)\ket = 0$ for an {\it 
arbitrary} operator ${\cal O}$.

Since the hamiltonian depends on the lapse ($\a$) and the shift ($\b$)  
functions, the time development of any quantity is generically gauge 
dependent, while the WDW pair of equations is gauge independent, because 
of the closure of the constraint algebra.
Equation~(\ref{Heisenberg}) may be interpreted in two ways.  First it is the 
ordinary classical canonical equation of motion for the field variables 
(obtained by replacing an operator with the corresponding canonical variable). 
In the present formulation of quantum gravity, it should be regarded as 
determining orbits of the operators in the Hilbert space.
The notion of the hamiltonian is redundant in  genuine quantum theory, but the 
form of the hamiltonian is consistently determined when one wishes to compare 
the quantum theory with the classical theory.

Since the hamiltonian is not Hermitian, an Hermitian operator generally 
becomes  a non-Hermitian operator by time development.
We postulate that the expectation value of an operator developed from an 
Hermitian operator, evaluated in a physical state, should be real.
This requirement restricts the gauge.
In what follows we show that the above condition leads to the gauge choice 
corresponding to the correct classical metric $g_{00}$ in the Schwarzschild 
coordinates.

The commutation relations are written, in the gauge $\b = 0$, as
\def\df{\fr{\d}{\d\vf(r)}}
\def\dd{\fr{\del}{\del r}}
\def\cD{{\cal D}}
\def\com{{\cal C}}
\def\NN{{\cal N}}
\be
	i{[}H, \r{]}  &=&  \a\vf'^{-1}e^{\r + \vf}\left[-\left(\NN' 
+ \NN\r'\right) + Ge^{\vf}\F_1\right], \la{Hr}\\
	i{[}H, \vf{]}  &=&  -\a e^{\r + \vf}\NN \la{Hf},
\ee
where $\NN  \equiv  Ge^{-2\r + \vf}i\dr$.  This quantity satisfies
\be
  i{[}H, \NN{]}  &=&  e^{-2\r + \vf}(\a  +  \a_{\r})\F_2 - G^{-1}e^{-3\r 
- \vf}(\a'\vf' + (\vf'^2  -  \vf^{[2]})\a) \la{HNN}.
\ee
If we choose
\be
	\a = \vf'e^{-(\r + \vf)} \la{gauge},
\ee
then the r.l.s. of Eq.~(\ref{HNN}) vanishes. In this gauge we have
\be
	\r(r, t) &=& \r(r) + \sum_{n=1}^{\infty}\fr{t^n}{n!}\left[a_n + 
\sum_{k = 1}^{n}b_{nk}\fr{\del^k\r}{\del r^k}\right] + (\F) \la{Rt}, \\
	\vf(r, t) &=& \vf(r) + \sum_{n=1}^{\infty}\fr{t^n}{n!}
\sum_{k = 1}^{n}b_{nk}\fr{\del^k\vf}{\del r^k},
\ee
where the coefficients $a_n$ and $b_{nk}$ are functions of $\NN$ and its 
derivatives, determined iteratively as $a_1 = -\NN', ~a_{n+1} = -(\NN a_n)'$ 
and $b_{11} = -\NN, ~b_{n+1 k} = -(b'_{nk} + b_{n k-1})\NN, ~b_{m0} = 
b_{m m+1} = 0$.
In Eq.~(\ref{Rt}) the term $(\F)$ represents terms which vanish if they act 
on the physical state satisfying $\F_1|\ket = \F_2|\ket = 0$.
Thus we see the expectation values of $\r(r, t)$ and $\vf(r, t)$ in the 
physical states are real. The gauge (\ref{gauge}) corresponds to the correct classical value of the metric, {i.e.}, $g_{00} = -\bra\a^2\ket = -\g_c^{-1}$ 
for $e^{-2\vf} = r^2$.
Note again that there exists no condition restricting $\a$  if the 
hamiltonian is Hermitian.


Some comments on the {\it Heisenberg} relation, Eq.~(\ref{Heisenberg}), are 
in order.
There are two  differences between  the present situation and the ordinary 
transformation theory of quantum mechanics; namely, in our case, the 
hamiltonian vanishes in the physical Hilbert space, and the hamiltonian is 
not Hermitian with respect to the inner product defined by Eq.~(\ref{inner1}).
{}From these facts the wave functional in the {\it Schr\"odinger 
representations}, denoted by $\Ps_S$, is divided into bra and ket functionals 
satisfying
\be
        i\fr{\del}{\del t}\Ps^{({\rm bra})}_S &=& H^{\dag}\Ps^{({\rm bra})}_S, 
\la{bra-eq}\\
        i\fr{\del}{\del t}\Ps^{({\rm ket})}_S &=& 0,
\la{ket-eq}
\ee
where $H^{\dag}$ is the Hermite conjugate of the hamiltonian, the existence 
of which is implied by the Riesz theorem.
Matrix elements of an arbitrary operator in the Schr\"odinger representation 
are the same as those in the Heisenberg representation due to Eqs.
~(\ref{bra-eq}) and (\ref{ket-eq}).\vss

I am grateful to C.-B. Kim and to the members of the Doyou seminar for 
discussions.


\end{document}